\documentclass[useAMS,usenatbib]{mn2e}

\usepackage{graphicx,amssymb,color}
\usepackage[normalem]{ulem}

\newcommand{\rev}{ }

\title[Constraining debris around ZTF J0139+5245]
{Constraining the origin of the planetary debris surrounding ZTF J0139+5245 through rotational fission
of a triaxial asteroid}

\author[Veras, McDonald \& Makarov]
{Dimitri Veras$^{1,2}$\thanks{E-mail: d.veras@warwick.ac.uk}\thanks{STFC Ernest Rutherford Fellow},
Catriona H. McDonald$^{1,2}$, Valeri V. Makarov$^3$
\\
$^{1}$Centre for Exoplanets and Habitability, University of Warwick, Coventry CV4 7AL, UK
\\
$^{2}$Department of Physics, University of Warwick, Coventry CV4 7AL, UK
\\
$^{3}$U.S. Naval Observatory, 3450 Massachusetts Ave., Washington, DC 20392-5420, USA
}

\pubyear{2019}

\begin{document}
\label{firstpage}
\pagerange{\pageref{firstpage}--\pageref{lastpage}}
\maketitle

\begin{abstract}
White dwarfs containing orbiting planetesimals or their debris represent crucial benchmarks by which theoretical investigations of post-main-sequence planetary systems may be calibrated. The photometric transit signatures of likely planetary debris in the ZTF J0139+5245 white dwarf system has an orbital period of about 110 days. An asteroid which breaks up to produce this debris may spin itself to destruction through repeated close encounters with the star {\it without} entering its Roche radius and without influence from the white dwarf's luminosity. Here, we place coupled constraints on the orbital pericentre ($q$) and the ratio ($\beta$) of the middle to longest semiaxes of a triaxial asteroid which disrupts outside of this white dwarf's Roche radius ($r_{\rm Roche}$) soon after attaining its 110-day orbit. We find that disruption within tens of years is likely when $\beta \lesssim 0.6$ and $q\approx 1.0-2.0r_{\rm Roche}$, and when $\beta \lesssim 0.2$ out to $q\approx 2.5r_{\rm Roche}$. Analysing the longer-timescale disruption of triaxial asteroids around ZTF J0139+5245 is desirable but may require either an analytical approach relying on ergodic theory or novel numerical techniques.  
\end{abstract}


\section{Introduction}

White dwarf planetary science is entering a new era with an emergent population of orbiting asteroids and their debris. The first three white dwarfs around which intact or currently disintegrating planetesimals have been discovered \citep{vanetal2015,manetal2019,vanetal2019} all present different and enticing challenges to the canonical dynamical model of the post-main-sequence evolution of minor planets.

As a star leaves the main sequence, its luminosity increases sufficiently to spin up minor planets to the point of rotational fission through the YORP effect \citep{veretal2014a,versch2020}. Although the extent and location of the break-up depends on the physical parameters of the asteroids in question, \cite{veretal2014a} claimed that asteroids under about 10 km in radius within about 7 au of their parent star are easily destroyed. Larger asteroids may orbitally migrate through another radiative effect known as the Yarkovsky effect \citep{veretal2015a,veretal2019}. The result is likely a sea of debris plus intact bodies in an annulus spreading from a few au to hundreds of au.

After the star has become a white dwarf, major planets such as the recently discovered WD J0914+1914b \citep{ganetal2019} could perturb these minor planets into the star's Roche radius, where the subsequent destruction could potentially be observable. This idea was pioneered by \cite{graetal1990} and \cite{jura2003}, and detailed numerical investigations of the breakup of these minor planets and the subsequent formation of discs were later conducted by \cite{debetal2012}, \cite{veretal2014b}, and \cite{malper2020a,malper2020b}. Following the first detection of a planetary debris disc around a white dwarf \citep{zucbec1987}, nearly three decades of observations have revealed over 40 such discs \citep{farihi2016} and over 1000 white dwarf atmospheres containing planetary remnants \citep{zucetal2003,zucetal2010,koeetal2014,couetal2019} before, finally, signatures of individual orbiting asteroids were discovered around WD 1145+017 \citep{vanetal2015}. 

Concurrent with these mounting observations, theories about the delivery of minor planets into the close vicinity of white dwarfs have matured \citep{veras2016}. The prevailing model is that any minor planet which reaches the vicinity of the white dwarf previously resided in an orbit with an eccentricity near unity and a semimajor axis of at least several au. In contrast, the photometric transit signatures of material surrounding WD 1145+017 indicate minor planets with orbital periods which are well constrained to be about 4.5 hours (corresponding to a semimajor axis $a$ of about $1R_{\odot} \approx~0.005$ au). Further, theoretical modelling has indicated that these asteroids are on nearly circular orbits, with eccentricities $e$ less than about 0.01 \citep{guretal2017,veretal2017,duvetal2020}.

In 2019, two more planetary systems were reported to contain inferred minor planet orbits which are at 
odds with the canonical model. \cite{manetal2019} detected spectroscopic signatures in SDSS J1228+1040 
which are indicative of a minor planet on a 2-hour orbit ($a = 0.73R_{\odot} \approx~0.0034$ au) with $e=0.54$. \cite{vanetal2019} then reported transit dips which are likely due to planetary debris orbiting the white dwarf ZTF J013906.17+524536.89 (ZTF J0139+5245). These dips indicate a $110$-day orbital period, suggesting $a=0.36$ au.

In order to be in agreement with the canonical model, the surrounding material would require a highly eccentric orbit in order for the progenitor to have entered the Roche radius and broken up. However, \cite{makver2019} recently demonstrated that a highly-eccentric triaxial asteroid may spin itself to the point of rotational fission {\it without} passing through the Roche radius and without any YORP-based contribution from the white dwarf's luminosity. The exchange of spin and orbital angular momentum during close encounters produces a chaotically-evolving spin evolution which allows this mechanism to operate. This result provides a novel pathway to breakup, one which relaxes the restrictive eccentricity constraint.

In this paper, we explore the possibility that the debris surrounding ZTF J0139+5245 originated from a progenitor asteroid which broke up outside of the white dwarf's Roche radius. In this scenario, and by assuming a realistic and computationally feasible breakup timescale, we can place constraints on the aspect ratios and density of the triaxial progenitor asteroid, coupled with the maximum pericentre distance at which breakup could have occurred. We set up the evolution in Section 2, perform simulations in Section 3 and summarize in Section 4.

\section{YORP-Less Rotational Fission}

In order for a triaxial asteroid to spin itself up to the point of fission without a luminosity boost from the central star, significant energy exchange must take place between the asteroid's spin and orbital angular momentum. In this section, we briefly describe this phenomenon.

Consider a homogeneous triaxial asteroid with density $\rho$ and semiaxes, in decreasing size order, of $\mathfrak{a}, \mathfrak{b}$, and $\mathfrak{c}$.  By utilising the following aspect ratios 

\begin{equation}
\alpha \equiv \frac{\mathfrak{c}}{\mathfrak{a}},
\end{equation}

\begin{equation}
\beta \equiv \frac{\mathfrak{b}}{\mathfrak{a}}, 
\end{equation}

\noindent{}we can remove one degree of freedom from the absolute sizes and treat the evolution as scale-free.
One particularly useful manifestation of these aspect ratios is in the equation of rotational
motion of the asteroid, which is independent of the asteroid's size and instead is a function
of $\beta$ \citep{danby1962,golpea1968}:

\begin{equation}
\frac{d^2\theta(t)}{dt^2} + Y(t) =0
,
\label{rotmot}
\end{equation}

\begin{equation}
\ \ \ \ \ \ \ \ \ \ \ \ \  Y(t) \equiv \frac{3}{2}n^2\left(\frac{1-\beta^2}{1+\beta^2} \right)
\frac{\sin{\left[2\theta(t) - 2 f(t)\right]}}{\left(1 - e \cos{E(t)}  \right)^3}
.
\label{rotmot2}
\end{equation}

\noindent{}Here, $\theta$ is the rotation angle about the shortest axis of the asteroid, $n$ represents its mean motion, $f$ represents its true anomaly and $E$ represents its eccentric anomaly.

The parameters $n$, $\beta$ and $e$ all remain constant throughout the evolution; the orbit of the asteroid is not assumed to change. Further, $n$ is a given value for ZTF J0139+5245 because the mass of the white dwarf is $M_{\star} =0.52M_{\odot}$, and a semimajor axis of $a = 0.36$ au then yields an orbital period of $P = 110$~days (consistent with the transit dips). When $\beta =1$, then $Y(t)=0$, and the asteroid is by definition oblate, and will not experience rotational fission. 

\cite{makver2019} integrated equation (\ref{rotmot}) for a few specific cases to indicate that when $Y(t)$ is sufficiently high, the time evolution of $d\theta/dt$ may exceed the breakup speed of the asteroid. We can determine how $Y(t)$ depends on key variables through

\begin{equation}
{\rm max}\left[Y(t)\right] = 
\frac{3}{2} v_{q}^2\left(\frac{1-\beta^2}{1+\beta^2}\right)
\propto q^{-3} \left(\frac{1-\beta^2}{1+\beta^2}\right)
\label{Yt}
\end{equation}

\noindent{}where $v_q$ is the speed at orbital pericentre. Equation (\ref{Yt}) illustrates explicitly that break up is more likely for lower $q$ and lower $\beta$. 

The functional dependencies in equation (\ref{Yt}), however, should be considered concurrently with the fact that $Y(t)$ is time dependent. The resulting chaotic nature of the asteroid's spin has been known for decades, and is strongly eccentricity-dependent: within Poincar\'{e} sections, the higher the eccentricity, the smaller the ``islands'' of regular trajectories around resonances and zones of circulation immersed in the chaotic ``swamp'' \citep{wisetal1984,wisdom1987}. When $e \approx 1$, the chaotic swamp dominates. \cite{makver2019} demonstrated that the $e \approx 1$ case requires considerable care in the numerical implementation because of the fluctuations in the denominator. This computational restriction highlights the potential benefit of the application of a predictive analytical model such as ergodic theory \citep[e.g.][]{stolei2019} to the long-term evolution of these systems.

The asteroid's breakup speed corresponds to the moment its spin reaches a critical limit $\omega_{\rm crit}$.
\cite{holsapple2007} obtained an explicit expression for the spin barrier limit in the strength regime (their Eqs. 5.9--5.10), when $\bar{R} = \left(\mathfrak{a}\mathfrak{b}\mathfrak{c}\right)^{1/3} \lesssim 10$ km.  We derive a more general, albeit slightly cumbersome, explicit expression that is applicable for asteroids of all sizes (encompassing both the strength and gravity regimes) by applying the expressions for the average stresses from \cite{holsapple2004} to the standard Drucker-Prager criterion\footnote{{\rev The Mohr-Coulomb failure criterion provides an alternative to the Drucker-Prager criterion used here. The former gives the maximum shear stress any plane can withstand as a function of the maximum and minimum principal stresses, resulting in six possible regimes with different orderings of the principal stresses. Instead, the Drucker-Prager criterion uses the square root of the second invariant of the deviator stress and includes a single relation for all stress states, removing this complication.  Both \cite{holmic2006} and \cite{wojciechowski2018} illustrated that both criteria give similar results except in a narrow region of phase space.}}. The result is

\[
\omega_{\rm crit} = \omega_{\rm crit}\left(\rho, \mathfrak{a}, \mathfrak{b}, \mathfrak{c}, \kappa, \phi \right)
\]
\begin{equation}
\ \ \ \ \ \ \,  = \sqrt{C_1 \left[C_2 + C_3 - \sqrt{C_4 + C_5 + C_6\left(C_7 + C_8\right)}  \right]}
,
\label{critrate}
\end{equation}

\noindent{}where the expressions for the $C$ variables are given in the Appendix, and $\kappa$ and $\phi$ refer to the strength coefficient and angle of friction respectively.

\section{Phase space exploration}

\begin{figure}
\includegraphics[width=8.8cm]{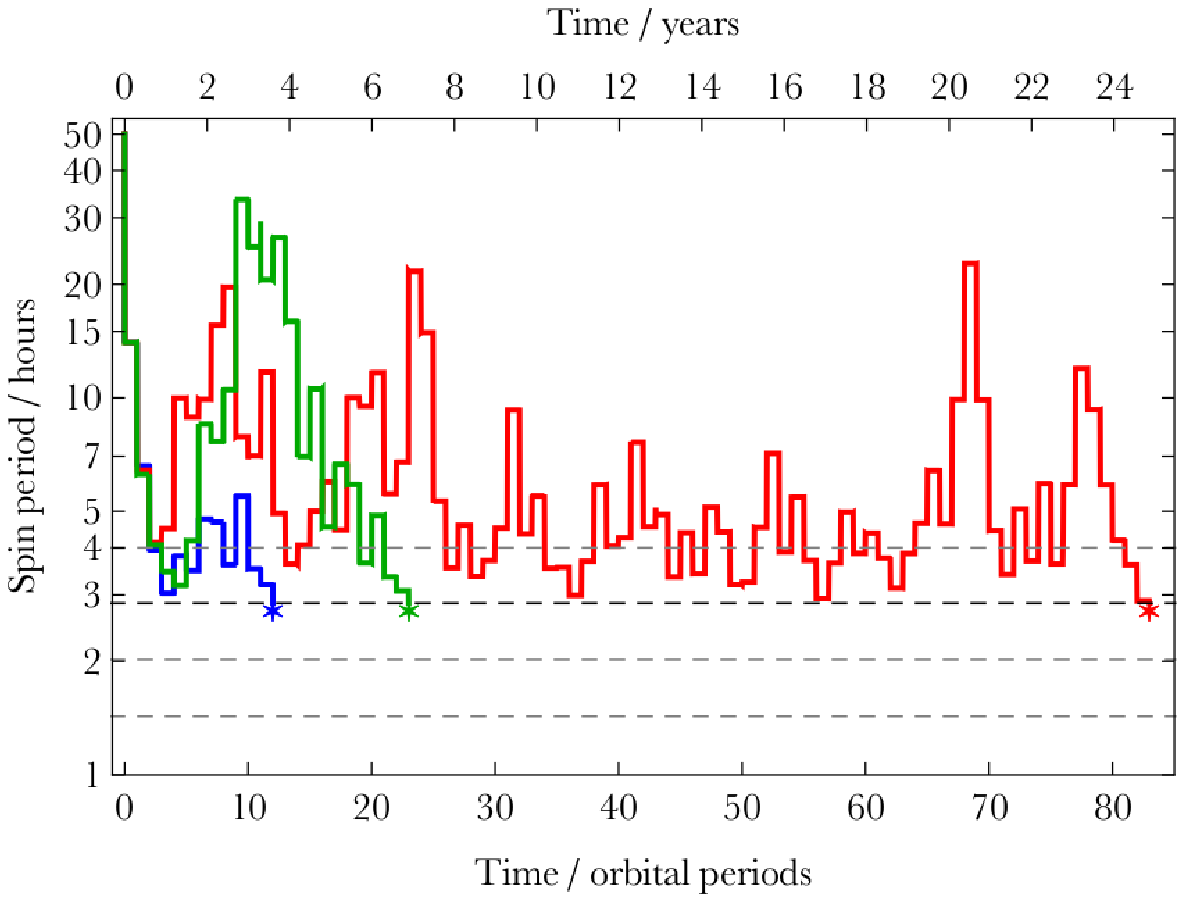}
\caption{
Three representative spin evolutions of triaxial asteroids with dimensions 
$\left\lbrace\mathfrak{a}, \mathfrak{b}, \mathfrak{c}\right\rbrace = \left\lbrace 100,75,55 \right\rbrace$ km orbiting ZTF J0139+5245 on a 110-day orbit with $q=1.3r_{\rm Roche}$ and an initial spin rate of one revolution every 50 hours. They all have $\rho = 2$~g/cm$^3$ and break up (at the locations of the six-pointed stars) at different times upon reaching the critical spin period (black dashed line at 2.85 hours). For perspective, gray dashed lines are shown corresponding to what the critical spin rate would have been for (from top to bottom), $\rho = 1,4,8$ g/cm$^3$. The only difference in the initial conditions for the simulations is a $3 \times 10^{-3}$ degree offset in their initial orientations, demonstrating the stochasticity of the evolution.
}
\label{Rep}
\end{figure}

We now perform numerical simulations; for details of the numerical challenges of integrating equation (\ref{rotmot}), see \cite{makver2019}.  In order to illustrate the stochastic nature of equation (\ref{rotmot}), we plot three representative evolution examples in Fig. \ref{Rep} with near-identical initial conditions. In each case, we set $\left\lbrace\mathfrak{a},\mathfrak{b},\mathfrak{c}\right\rbrace = \left\lbrace 100, 75, 55 \right\rbrace$ km, $\rho = 2$ g/cm$^3$, the initial spin rate to one revolution every 50 hours (i.e. almost stationary), and $q=1.3r_{\rm Roche}$. The only difference in the initial conditions of the three simulations is the value of $\theta(0)$, which is taken to be $5.497^{\circ}$ (blue curve), $5.500^{\circ}$ (red curve) and $5.503^{\circ}$ (green curve).

Here, the Roche radius ($r_{\rm Roche}$) is defined by assuming {\rev that the asteroid is a solid, spinning rubble pile} through \citep{veretal2017}:

\begin{equation}
r_{\rm Roche} = 0.94 R_{\odot} \left( \frac{0.52 M_{\odot}}{0.60M_{\odot}} \right)^{1/3} 
                          \left(\frac{\rho}{3 \ {\rm g/cm}^3} \right)^{-1/3}
.
\label{rroche}
\end{equation}

{\rev \noindent{}Because the spin of the asteroid changes with each periastron passage, the instantaneous Roche radius is actually a function of time. Also, non-rubble pile asteroids have non-zero internal strength, which would change the Roche radius \citep{beasok2015}. Hence, equation~(\ref{rroche}) represents an upper bound with respect to this parameter; incorporating nonzero strength can only decrease, but not increase, the Roche radius from its currently given value.}

Figure \ref{Rep} illustrates that the asteroids reach the critical spin period (black dashed line, at 2.85 hours) after respectively 12, 83 and 23 orbits (denoted by six-pointed stars). Other dashed lines (all gray) are also displayed to indicate what would have been the critical spin limit if $\rho$ was different. The four dashed lines from top to bottom correspond to $\rho = \left\lbrace 1,2,4,8 \right\rbrace$ g/cm$^3$. The dependence of $\omega_{\rm crit}$ on the aspect ratios of the asteroids is smaller. For example, for $\mathfrak{a} = 100$ km, any values of $\mathfrak{b}$ and $\mathfrak{c}$ greater than 55 km yield a critical spin period between 2.56 and 3.02 hours. In a more extreme case, however, when $\mathfrak{b} =\mathfrak{c} = $ 20 km, then the critical period becomes 5.49 hours.

For our more general phase space exploration, we focus on the two key variables $\beta$ and $q$. Because $Y(t)$ is independent of $\alpha$ and is scale-free, we set $\alpha = \beta$ (the prolate case) and arbitrarily choose $\mathfrak{a} = 100$ km. Although varying $\alpha$ and $\mathfrak{a}$ would alter $\omega_{\rm crit}$, this variation would not be sufficiently high to qualitatively change our results. We also set the initial value of $d \theta / dt$ to one revolution every 250 hours, roughly two orders of magnitude higher than the critical spin limit for spherical asteroids, to mimic an initially stationary asteroid. We choose the value of $\theta(0)$ randomly from a uniform distribution for each simulation.

We split our simulation into three sets for different values of $\rho$. Doing so changes $r_{\rm Roche}$, and hence the absolute scale of the values of $q$ that we sample. For example, for $\rho = 1, 4, 8$ g/cm$^3$, $r_{\rm Roche} = 1.29, 0.81, 0.65R_{\odot}$. We note that the $\rho = 8$ g/cm$^3$ case is important to consider because the planetesimal orbiting SDSS J1228+1040 \citep{manetal2019} is likely a dense planetary core fragment rather than a rubble pile asteroid. 

\begin{figure}
\centerline{
\includegraphics[width=8.5cm]{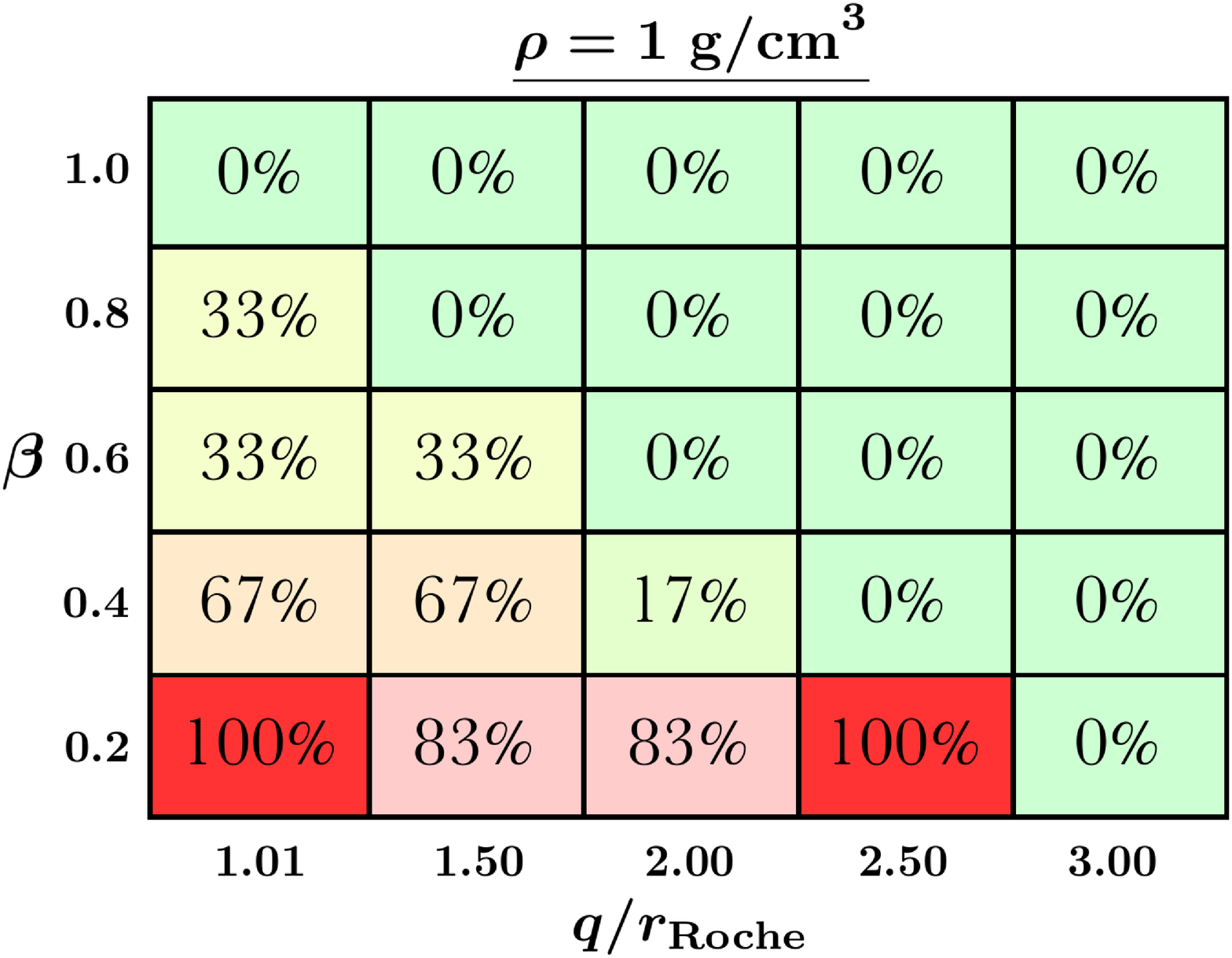}
}
\centerline{}
\centerline{
\includegraphics[width=8.5cm]{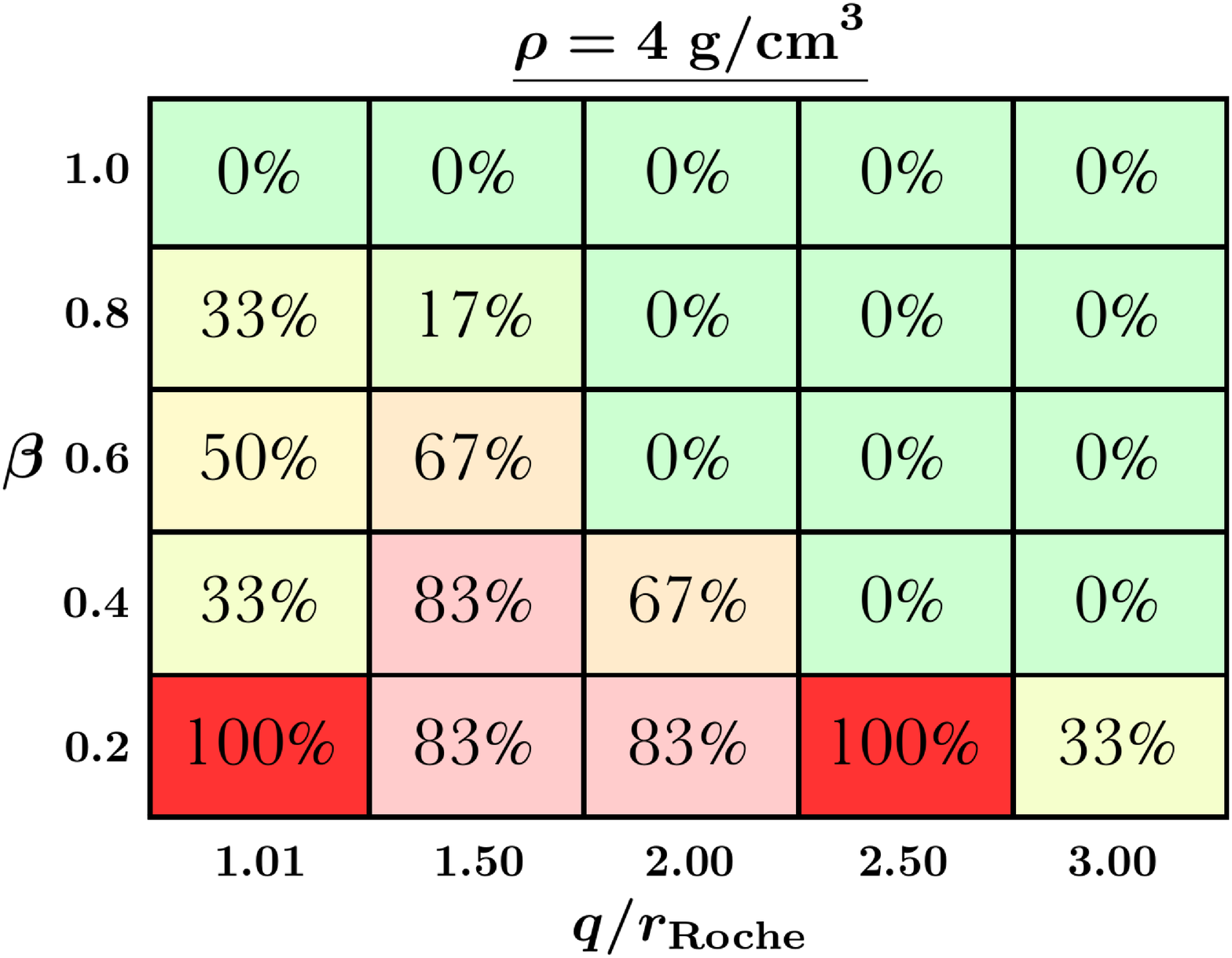}
}
\centerline{}
\centerline{
\includegraphics[width=8.5cm]{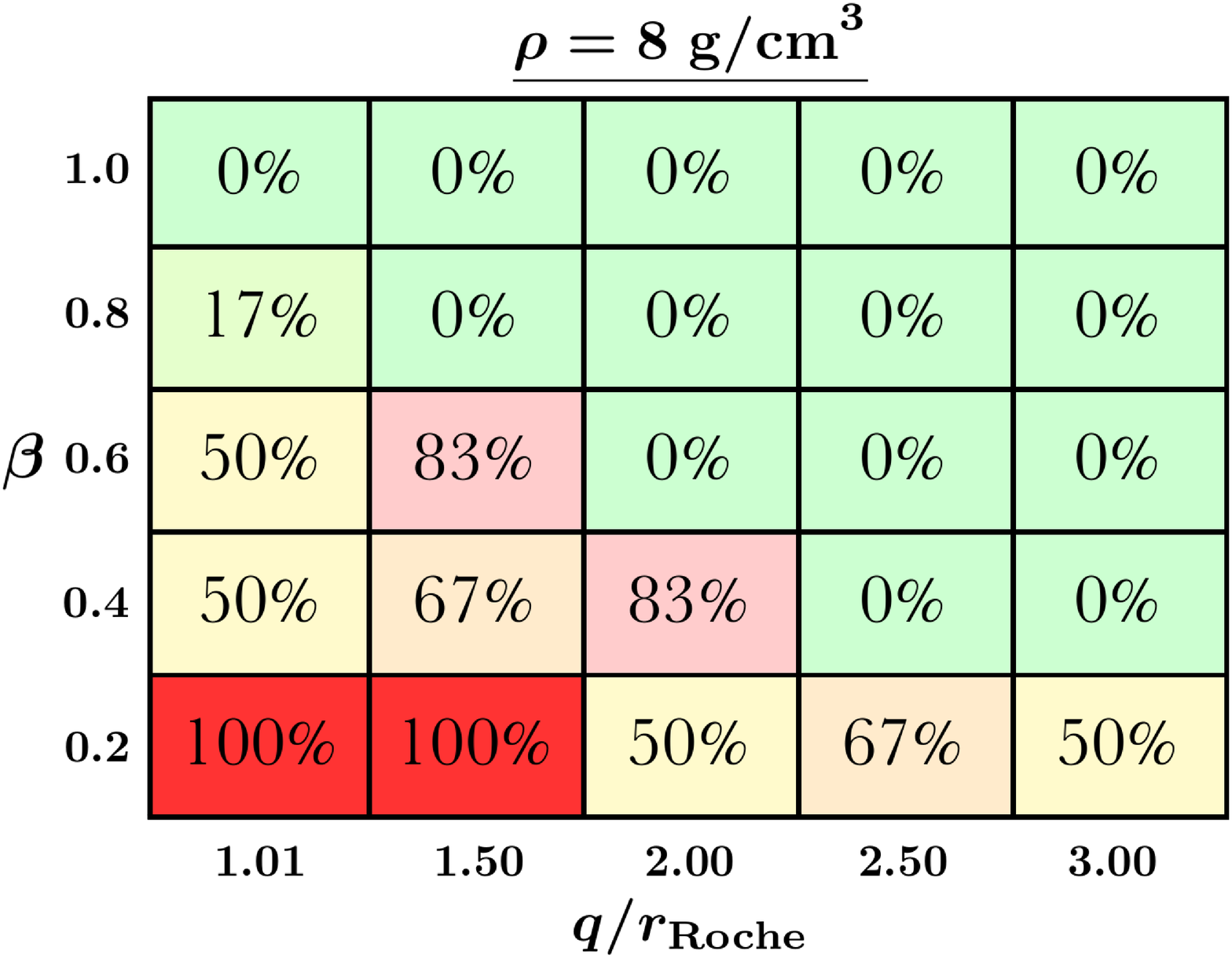}
}
\caption{
Break-up fractions as a function of the asteroid aspect ratio $\beta \equiv \mathfrak{b}/\mathfrak{a}$ and pericentre distance $q$ in terms of the Roche radius of ZTF J0139+5245 (equation \ref{rroche}). For each box, six simulations were performed, and the inset percentage indicates the fraction of simulations where the asteroid breaks apart.
}
\label{Tabs}
\end{figure}

For a given triplet of $\left(\rho, \beta, q\right)$, we run six simulations for 100 orbits each. That value corresponds to the upper limit of the number of orbits we can confidently propagate with our numerical implementation with $q$ values so close to the Roche radius. We report on the fraction of unstable simulations out of six within each box in Fig. \ref{Tabs}.

The figure demonstrates a clear trend of greater instability for decreasing $\beta$ and $q$, which is predicted from equation (\ref{Yt}). Over the timescale of only 100 orbits, breakup can occur out to $3r_{\rm Roche}$ for the most prolate asteroids. For asteroids which are more spherical ($\beta \lesssim 0.6$), breakup is common in the range $1-2r_{\rm Roche}$. These statements hold true across all possible asteroid densities. Although the figure displays clear trends, the tables showcase a nonuniformity which is indicative of the chaotic nature of the evolution.

\section{Discussion}

\cite{vanetal2019} suggest that the progenitor of the debris around ZTF J0139+5245 may be experiencing an early phase of tidal disruption. One way to distinguish an early or late phase of disruption is to determine whether a full ring of debris has formed, and whether these ring particles have started to sublimate or drift towards the white dwarf through Poynting-Robertson drag \citep{stoetal2015,veretal2015b,broetal2017}.

When an asteroid breaks up, the differential speed of the components allows the debris to eventually spread out into a ring. If these particles are not perturbed from their original orbit and are treated as collisionless, then the filling time $t_{\rm fill}$ of the ring may be analytically approximated. By assuming an initially spherical asteroid of $R$ and that breakup occurs instantaneously and at the orbital pericentre $q = a(1-e)$, \cite{veretal2017} derived this equation in their Equation 8. In the limit $e \rightarrow 1$, the expression becomes

\begin{equation}
\frac{t_{\rm fill}}{P} \approx \frac{a}{6R}\left(1-e\right)^2
.
\end{equation}

\noindent{}For aspherical asteroids like the ones we are considering here, we can make a further approximation and replace $R$ with $\bar{R}$ \citep{holsapple2007}. Then, because $P$ and $a$ are known quantities, we can express the filling time for a triaxial asteroid breakup around ZTF J0139+5245 as

\[
t_{\rm fill} \approx \frac{P}{6} \left(\frac{q}{\bar{R}}\right) \left(\frac{q}{a}\right)
\]
\begin{equation}
\ \ \ \ \ 
\approx 18.0 \ {\rm yr} 
\left( \frac{\bar{R}}{100 \ {\rm km}} \right)^{-1}
\left( \frac{q}{2 R_{\odot}} \right)^{2}.
\label{Tfill}
\end{equation}

Relative to the white dwarf's estimated cooling age (about 500 Myr; \citealt*{vanetal2019}), 18 yrs is virtually instantaneous. If a full ring has not yet formed from a disruption event, then either the assumptions which enter equation (\ref{Tfill}) are too simplistic, the rate of asteroid delivery close to the Roche radius is particularly high, the breakup of the asteroid occurred well beyond $3 r_{\rm Roche}$, and/or we are observing the system at a fortuitous time. Although the debris rings formed around WD 1145+017 break apart and re-form on yearly timescales (see Bruce Gary's observation log at http://www.brucegary.net/1145/ and for ring arcs, \citealt*{izqetal2018}), the entirety of those orbits are at or near the sublimation radius (coincidentally at an approximate distance of $1r_{\rm Roche}$) for a variety of materials \citep{rafgar2012}. In contrast, for ZTF J0139+5245, the pericentre of the ring may be at $2-3 r_{\rm Roche}$, outside of the sublimation radius.

We emphasize that the characteristic time for spin-up and rotational fission depends more on the progenitor's degree of elongation than on its average density. As observed in the Solar system, when an object's mass approaches that of a major planet, its elongation approaches zero. Such large bodies still experience chaotic spin evolution, but at a negligibly small level, too small to explain breakup around ZTF J0139+5245. 

We can corroborate this conclusion by estimating the mass of the progenitor of the debris. Although equations (\ref{rotmot}--\ref{rotmot2}) are independent of size, we can estimate the mass by considering the duration of the transits. If we assume that (i) each transit is equivalent to a total eclipse lasting four days, (ii) the dust resides in a rectangular cloud with height equal to the white dwarf's diameter, and (iii) individual dust grains of radius $R_{\rm dust}$ fill the cloud without overlapping, then the total mass of all dust grains $M_{\rm dust}$ is

\begin{equation}
M_{\rm dust} \approx \frac{4}{3} \pi^2 \left( \frac{4 \ {\rm days}}{110 \ {\rm days}}\right) R_{\rm WD} R_{\rm dust} \rho_{\rm dust} a. 
\end{equation}

\noindent{}Given $R_{\rm dust} = 1 \mu$m, $\rho_{\rm dust} = 2$ g/cm$^3$ and $a = 0.72$ au, we find $M_{\rm dust} \approx 6 \times 10^{17}$ g (a small asteroid), which corresponds with the estimate from \cite{vanetal2019}. {\rev According to the model of \cite{vanetal2018}, this dust mass is too small for second-generation formation of asteroids to be a viable possibility.}

{\rev Nevertheless, other models besides the one we have posed here to explain the origin of the debris may be viable. The standard Roche disruption scenario (where a major planet kicks an asteroid directly into the Roche radius) may still be valid for this system, and we do not claim otherwise. However, this canonical model encounters difficulties if the asteroid is near-spherical and settles on an orbit whose pericentre exceeds the Roche radius. Then it will never disrupt, and not even perturb itself into the Roche radius through sublimation \citep{veretal2015c}. Also, near-spherical asteroids are not the norm in our solar system. Another potential source of the debris is due to catastrophic collisions from giant impacts \citep[e.g.][]{kenbro2005,jacetal2014}. }

{\rev What is the post-fission fate of the particles? \cite{scheeres2018} and \cite{versch2020} demonstrated that multiple generations of radiation-induced rotational breakup of a nearly-spherical rubble pile can occur until the rubble pile has been broken down into its monolithic components. In contrast here, for a triaxial rubble pile around ZTF~J0139+5245, the situation is more complex: rotational breakups can occur repeatedly until the child particles become spherical (when $\beta = 1$ in equation \ref{rotmot2}) \footnote{{\rev Both the absolute sizes and the initial spin values which are assumed for the child particles would be irrelevant, because they can chaotically reach breakup speed from a stationary start, as shown in Fig. \ref{Rep}.}}.  Also, the sibling particles of a single generation would not necessarily all have the same $\beta$ values: as a result, the timescales for the next breakup to occur may differ significantly amongst these particles, or not occur at all. Modelling the resulting collisional evolution would require the use of a sophisticated numerical code (such as the cascade code used in \citealt*{kenbro2017a,kenbro2017b}) that is beyond the scope of this paper.}

{\rev Can we predict how future observations would change if the debris around ZTF~J0139+5245 arises from rotational breakup of a triaxial asteroid outside of the Roche radius? The most robust observational link to a particular model would most likely arise from detecting secure morphological changes in the photometric transit dips between epochs. Those signatures would most likely indicate on-going disruption rather than natural features arising from dust-gas interactions, because unlike the compact gaseous and dusty disc around WD 1145+017 \citep{vanetal2015}, what is observed orbiting ZTF J0139+5245 is most likely a singular, more extended annulus. If the on-going disruption occurs and is due to Roche disruption, then changes may be visible after each pericentre passage. If, instead, the disruption is due to rotational disruption, then the disruption is likely to be more infrequent and not visible for at least many pericentre passages.}

\section{Conclusion}

{\rev Independent} of how or when the progenitor of the debris orbiting (ZTF J0139+5245) reached a 110-day orbit, in this paper we showed that regardless of its size, as long as the asteroid was aspherical, it may have broken up well {\it outside} of the Roche radius. This assumption allowed us to place coupled constraints on the prolateness of the asteroid and its orbital pericentre (Fig. \ref{Tabs}), constraints we found to be largely independent of density despite that parameter's strong effect on the critical spin rate (equation \ref{critrate}). In addition to providing an explanation for the debris orbiting ZTF J0139+5245, our simulations also suggest that the canonical assumption of minor planet disruption within the Roche radius of white dwarfs may need to be revised.

\section*{Acknowledgements}

{\rev We thank the anonymous MNRAS referee} and the internal USNO referees Michael Efroimsky and Robert T. Zavala for particularly valuable comments which have improved the manuscript. DV gratefully acknowledges the support of the STFC via an Ernest Rutherford Fellowship (grant ST/P003850/1).

\appendix

\section{Expressions for critical spin rate}

Here we provide explicit expressions for the critical spin rate in equation (\ref{critrate}).
The $C$ variables are

\begin{equation}
C_1= \left[\rho \mathfrak{a}^4 \left(3 s^2 \left(1 + \beta^2\right)^2 - 1 + \beta^2 - \beta^4  \right)  \right]^{-1}
,
\end{equation}

\begin{equation}
C_2= 15 s k \mathfrak{a}^2 \left(1 + \beta^2 \right)
,
\end{equation}

\[
C_3= \pi G \rho^2 \mathfrak{a}^4 
\bigg[ \left(1 + 6 s^2\right) \left(\left(A_x + A_y + A_z \alpha^2 \right)  \beta^2 + A_z \alpha^2 \right) 
\]

\begin{equation}
\ \ \ \ - 2 \left(1 - 3s^2\right) \left(A_x + A_y \beta^4 \right) \bigg]
,
\end{equation}

\begin{equation}
C_4= 75 k^2 \mathfrak{a}^4 \left(1 - \beta^2 + \beta^4 \right)
,
\end{equation}

\[
C_5= 90 \pi G k s \rho^2 \mathfrak{a}^6 
\]

\begin{equation}
\ \ \ \ \,  \times \left[A_z \alpha \left(1 + \beta^4 \right) - \beta^2 \left(1 - \beta^2 \right) \left(A_x - A_y\right)   \right]
,
\end{equation}

\begin{equation}
C_6= 3 \pi^2 G^2 \rho^4 \mathfrak{a}^8
,
\end{equation}

\[
C_7= \beta^4 \left(A_x - A_y\right)^2 \left(12s^2 - 1\right)
\]

\begin{equation}
\ \ \ \ \, + 2 A_z \alpha^2 \beta^2 \left(\beta^2 - 1\right) \left(A_x - A_y\right) \left(1 + 6 s^2\right) 
,
\end{equation}

\begin{equation}
C_8= A_{Z}^2 \alpha^4 \left[12s^2 \left(1 + \beta^2 + \beta^4\right) - \left(1 - \beta^2  \right)^2 \right]
.
\end{equation}

\noindent{}These variables are a function of the slope constant $s$, which is related to the angle of friction $\phi$ (assumed to be $45^{\circ}$, \citealt*{holsapple2007}) through

\begin{equation}
s = \frac{2 \sin{\phi}}{\sqrt{3} \left(3 - \sin{\phi}\right)}
,
\end{equation}

\noindent{}the shear strength intercept $k$, which is related to the strength coefficient $\kappa = 2.26 \times 10^6$ N/m$^{3/2}$ from

\begin{equation}
k = \kappa \mathfrak{a}^{-1/2} \left(\alpha \beta \right)^{-1/6}
,
\end{equation}

\noindent{}and the following integrals

\begin{equation}
A_x = \alpha \beta \int_{0}^{\infty} \frac{du}
{\left(u+1\right)^{3/2}\left(u + \beta^2\right)^{1/2}\left(u + \alpha^2\right)^{1/2}}
,
\end{equation}

\begin{equation}
A_y = \alpha \beta \int_{0}^{\infty} \frac{du}
{\left(u+1\right)^{1/2}\left(u + \beta^2\right)^{3/2}\left(u + \alpha^2\right)^{1/2}}
,
\end{equation}

\begin{equation}
A_z = \alpha \beta \int_{0}^{\infty} \frac{du}
{\left(u+1\right)^{1/2}\left(u + \beta^2\right)^{1/2}\left(u + \alpha^2\right)^{3/2}}
.
\end{equation}

\label{lastpage}

\begin{thebibliography}{99}

\bibitem[Bear \& Soker(2015)]{beasok2015} Bear, E., \& Soker, N.\ 2015, MNRAS, 450, 4233

\bibitem[Brown et al.(2017)]{broetal2017} Brown, J.~C., Veras, D., \& G{\"a}nsicke, B.~T.\ 2017, MNRAS, 468, 1575 

\bibitem[Coutu et al.(2019)]{couetal2019} Coutu, S., Dufour, P., Bergeron, P., et al.\ 2019, Submitted to ApJ, arXiv:1907.05932

\bibitem[Danby(1962)]{danby1962} Danby, J.\ 1962, Fundamentals of Celestial Mechanics, New York: Macmillan

\bibitem[Debes et al.(2012)]{debetal2012} Debes, J.~H., Walsh, K.~J., \& Stark, C.\ 2012, ApJ, 747, 148 

\bibitem[Duvvuri et al.(2020)]{duvetal2020} Duvvuri, G., Redfield, S., \& Veras, D.\ 2020, Submitted to ApJ

\bibitem[Farihi(2016)]{farihi2016} Farihi, J.\ 2016, New Astronomy Reviews, 71, 9 

\bibitem[G{\"a}nsicke et al.(2019)]{ganetal2019} G{\"a}nsicke, B.~T., Schreiber, M.~R., Toloza, O., et al.\ 2019, Nature, 576, 61

\bibitem[Goldreich \& Peale(1968)]{golpea1968} Goldreich, P., \& Peale, S.~J.\ 1968, ARA\&A, 6, 287

\bibitem[Graham et al.(1990)]{graetal1990} Graham, J.~R., Matthews, K., Neugebauer, G., \& Soifer, B.~T.\ 1990, ApJ, 357, 216 

\bibitem[Gurri et al.(2017)]{guretal2017} Gurri, P., Veras, D., \& G{\"a}nsicke, B.~T.\ 2017, MNRAS, 464, 321 

\bibitem[Holsapple(2004)]{holsapple2004} Holsapple, K.~A.\ 2004, Icarus, 172, 272

\bibitem[Holsapple(2007)]{holsapple2007} Holsapple, K.~A.\ 2007, Icarus, 187, 500

\bibitem[Holsapple \& Michel(2006)]{holmic2006} Holsapple, K.~A., \& Michel, P.\ 2006, Icarus, 183, 331

\bibitem[Izquierdo et al.(2018)]{izqetal2018} Izquierdo, P., Rodr{\'\i}guez-Gil, P., G{\"a}nsicke, B.~T., et al.\ 2018, MNRAS, 481, 703.

\bibitem[Jackson et al.(2014)]{jacetal2014} Jackson, A.~P., Wyatt, M.~C., Bonsor, A., et al.\ 2014, MNRAS, 440, 3757

\bibitem[Jura(2003)]{jura2003} Jura, M.\ 2003, ApJL, 584, L91 

\bibitem[Kenyon \& Bromley(2005)]{kenbro2005} Kenyon, S.~J., \& Bromley, B.~C.\ 2005, AJ, 130, 269

\bibitem[Kenyon \& Bromley(2017a)]{kenbro2017a} Kenyon, S.~J., \& Bromley, B.~C.\ 2017a, ApJ, 844, 116

\bibitem[Kenyon \& Bromley(2017b)]{kenbro2017b} Kenyon, S.~J., \& Bromley, B.~C.\ 2017b, ApJ, 850, 50

\bibitem[Koester et al.(2014)]{koeetal2014} Koester, D., G{\"a}nsicke, B.~T., \& Farihi, J.\ 2014, A\&A, 566, A34 

\bibitem[Makarov \& Veras(2019)]{makver2019} Makarov, V.~V., \& Veras, D.\ 2019, In Press, AAS Journals, arXiv:1908.04612

\bibitem[Malamud \& Perets(2020a)]{malper2020a} Malamud, U., Perets, H.\ 2020a, Submitted to MNRAS, arXiv:1911.12068

\bibitem[Malamud \& Perets(2020b)]{malper2020b} Malamud, U., Perets, H.\ 2020b, Submitted to MNRAS, arXiv:1911.12184

\bibitem[Manser et al.(2019)]{manetal2019} Manser, C.~J., G{\"a}nsicke, B.~T., Eggl, S., et al.\ 2019, Science, 364, 66

\bibitem[Rafikov \& Garmilla(2012)]{rafgar2012} Rafikov, R.~R., \& Garmilla, J.~A.\ 2012, ApJ, 760, 123

\bibitem[Scheeres(2018)]{scheeres2018} Scheeres, D.~J.\ 2018, Icarus, 304, 183

\bibitem[Stone et al.(2015)]{stoetal2015} Stone, N., Metzger, B.~D., \& Loeb, A.\ 2015, MNRAS, 448, 188 

\bibitem[Stone \& Leigh(2019)]{stolei2019} Stone, N.~C., \& Leigh, N.~W.~C.\ 2019, arXiv:1909.05272

\bibitem[van Lieshout et al.(2018)]{vanetal2018} van Lieshout, R., Kral, Q., Charnoz, S., et al.\ 2018, MNRAS, 480, 2784

\bibitem[Vanderbosch et al.(2019)]{vanetal2019} Vanderbosch, Z., Hermes, J.~J., Dennihy, E., et al.\ 2019, Submitted to ApJL, arXiv:1908.09839

\bibitem[Vanderburg et al.(2015)]{vanetal2015} Vanderburg, A., Johnson, J.~A., Rappaport, S., et al.\ 2015, Nature, 526, 546 

\bibitem[Veras et al.(2014a)]{veretal2014a} Veras, D., Jacobson, S.~A., G\"{a}nsicke, B.~T.\ 2014a, MNRAS, 445, 2794 

\bibitem[Veras et al.(2014b)]{veretal2014b} Veras, D., Leinhardt, Z.~M., Bonsor, A., G\"{a}nsicke, B.~T.\ 2014b, MNRAS, 445, 2244

\bibitem[Veras et al.(2015a)]{veretal2015a} Veras, D., Eggl, S., G{\"a}nsicke, B.~T.\ 2015a, MNRAS, 451, 2814 

\bibitem[Veras et al.(2015b)]{veretal2015b} Veras, D., Leinhardt, Z.~M., Eggl, S., G{\"a}nsicke, B.~T.\ 2015b, MNRAS, 451, 3453 

\bibitem[Veras et al.(2015c)]{veretal2015c} Veras, D., Eggl, S., \& G{\"a}nsicke, B.~T.\ 2015c, MNRAS, 452, 1945

\bibitem[Veras(2016)]{veras2016} Veras, D.\ 2016, Royal Society Open Science, 3, 150571 

\bibitem[Veras et al.(2017)]{veretal2017} Veras, D., Carter, P.~J., Leinhardt, Z.~M., \& G{\"a}nsicke, B.~T.\ 2017, MNRAS, 465, 1008 

\bibitem[Veras et al.(2019)]{veretal2019} Veras, D., Higuchi, A., \& Ida, S.\ 2019, MNRAS, 485, 708

\bibitem[Veras \& Scheeres(2020)]{versch2020} Veras, D. \& Scheeres, D.~J.\ 2020, MNRAS In Press, arXiv:2001.00949

\bibitem[Wisdom et al.(1984)]{wisetal1984} Wisdom, J., Peale, S.~J., \& Mignard, F.\ 1984, Icarus, 58, 137

\bibitem[Wisdom(1987)]{wisdom1987} Wisdom, J.\ 1987, AJ, 94, 1350

\bibitem[Wojciechowski(2018)]{wojciechowski2018} Wojciechowski, M.\ 2018, Studia Geotechnica et Mechanica, 40, 163

\bibitem[Zuckerman \& Becklin(1987)]{zucbec1987} Zuckerman, B., \& Becklin, E.~E.\ 1987, Nature, 330, 138

\bibitem[Zuckerman et al.(2003)]{zucetal2003} Zuckerman, B., Koester, D., Reid, I.~N., H\"{u}nsch, M.\ 2003, ApJ, 596, 477 

\bibitem[Zuckerman et al.(2010)]{zucetal2010} Zuckerman, B., Melis, C., Klein, B., Koester, D., \& Jura, M.\ 2010, ApJ, 722, 725 

\end{thebibliography}
\end{document}